\documentclass{article}
\usepackage{graphicx} 
\usepackage{amsmath}

\usepackage{natbib}
\newcommand{\dbar}{d\hspace*{-0.08em}\bar{}\hspace*{0.1em}}

\usepackage{soul}
\usepackage{color}
\renewcommand{\hl}[1]{#1}

\usepackage{footmisc}
\usepackage{tabularx}
\usepackage{array}

\usepackage{enumerate}

\usepackage{multirow}

\usepackage{MnSymbol}

\title{Making Sense of Gravitational Thermodynamics}
\author{Lorenzo Lorenzetti\footnote{Institute of Philosophy, Università della Svizzera italiana, Lugano, Switzerland. Contact: lorenzo.lorenzetti.ac@gmail.com}}
\date{(Forthcoming in \textit{Philosophy of Physics})}

\begin{document}

\maketitle

\begin{abstract}

The use of statistical methods to model gravitational systems is crucial to physics practice, but the extent to which thermodynamics and statistical mechanics genuinely apply to these systems is a contentious issue. This paper provides new conceptual foundations for gravitational thermodynamics by reconsidering the nature of key concepts like equilibrium and advancing a novel way of understanding thermodynamics. The challenges arise from the peculiar characteristics of the gravitational potential, leading to non-extensive energy and entropy, negative heat capacity, and a lack of standard equilibrium. Hence it has been claimed that only non-equilibrium statistical mechanics is warranted in this domain, whereas thermodynamics is inapplicable. We argue instead that equilibrium statistical mechanics applies to self-gravitating systems at the relevant scale, as they display equilibrium in the form of metastable quasi-equilibrium states. We then develop a minimal framework for thermodynamics that can be applied to these systems and beyond. Thermodynamics applies in the sense that we can devise macroscopic descriptions and explanations of the behaviour of these systems in terms of coarse-grained quantities like energy and temperature within equilibrium statistical mechanics.

\end{abstract}

\tableofcontents

\section{Introduction}

While possible in theory, tracking the evolution of self-gravitating systems (SGS) like star clusters by calculating the trajectories of a million stars at a time is practically impossible. Hence the application of statistical methods is crucial. However, the precise extent to which statistical mechanics and thermodynamics genuinely apply to these systems is a highly contentious matter. 

This topic is an important and foundationally interesting area of physics that has been understudied and is ripe for philosophical work. It also provides an interesting case study that will motivate us to advance a novel way to understand thermodynamics with further possible applications.

The challenges arise due to the peculiar long-range nature of gravity in these systems. The gravitational potential distinguishes them from more conventional short-range interacting systems. It entails unconventional thermodynamic and statistical mechanical behaviour, especially non-extensivity of energy and entropy, negative heat capacity, and lack of standard equilibrium.\footnote{See Padmanabhan (\citeyear{padmanabhan1989antonov}, \citeyear{padmanabhan1990statistical}, \citeyear{padmanabhan2008statistical}), \cite{chavanis2002thermodynamics}, \cite{Katz2003-KATTOS}, \cite{heggie_hut_2003}, \cite{binney2011galactic}, \cite{sormani2013gravothermal}, \cite{campa2014physics}.} 

In the limited debate on the issue, some have maintained that thermodynamics could still be suitable to describe these systems, provided we revise certain thermodynamic features that are ascribed to conventional systems. In contrast, others have claimed that thermodynamics is unfit to model these systems and that only the application of non-equilibrium statistical mechanics is supported.\footnote{\cite{Callender2011-CALHAH} is the main figure in the first camp and defends a view that is arguably supported by physicists such as \cite{lynden1968gravo}, \cite{chavanis2002thermodynamics}, \cite{Katz2003-KATTOS}, \cite{heggie_hut_2003}. \cite{Robertson2019-ROBSAS-18} defends explicitly the second view, voicing the scepticism by e.g. \cite{ruelle1969statistical}, \cite{binney2011galactic}. See also \cite{wallace2010gravity}.} 

This paper advances our understanding of gravitational physics in two ways. First, we argue that \textit{equilibrium} statistical mechanics can be meaningfully applied to SGS in the appropriate regime, alongside non-equilibrium statistical mechanics. This is supported by the fact that equilibrium can be found in the form of metastable quasi-equilibrium states in these systems at certain scales, and by the idea that equilibrium is essentially scale-relative.\footnote{Metastable quasi-equilibrium refers to local equilibria which are relatively stable under perturbations and effectively in equilibrium over a certain time scale.} We prove this point first within idealised models and then within more realistic models of globular clusters, applying Earman's (\citeyear{earman2004curie}) principle for de-idealisation.

Second, although \hl{full-blown phenomenological thermodynamic descriptions} are unsuitable in this domain, we develop what we call a `minimal framework for thermodynamics' and show how a notion of thermodynamics applies to SGS. In fact, we can provide thermodynamic explanations based on the behaviour of macro-level quantities like temperature and energy describing these systems within the domain of equilibrium statistical mechanics. While non-equilibrium and equilibrium statistical mechanics and phenomenological thermodynamics are distinct theories, minimal thermodynamics is a coarse-grained \textit{level of description} within the framework of equilibrium statistical mechanics. It qualifies as `thermodynamics' especially in virtue of its use of macroscopic coarse-grained quantities partially autonomous from the microscopic variables characterising purely statistical descriptions. We maintain that the picture we develop is the best way to make sense of the notion of gravitational thermodynamics.

The paper has two goals. First, it provides new conceptual foundations for the study of gravitational thermodynamics by drawing a clear map of statistical and thermal physics and elucidating how they apply to the physics of SGS. In particular, we provide a more accurate reconstruction of physics practice in this field by showing how certain phenomena like core collapse in globular clusters can be predicted and explained starting both from distinctively statistical mechanical and coarser-grained thermodynamic points of view. Despite the unconventional features of these systems, these different methodologies are all equally justified at the right scale, while there are natural trade-offs between more complex but richer statistical mechanical explanations and simpler but more limited thermodynamic explanations of the same phenomena. Maintaining that only non-equilibrium statistical mechanics applies to these systems misses crucial aspects of SGS physics which fall within equilibrium statistical mechanics and (minimal) thermodynamics.

Second, analysing SGS allows us to draw general lessons about thermodynamics and statistical mechanics and to study the impact of idealisations in astrophysical models. The case study supports a more liberal approach to concepts such as equilibrium \citep{Callender2001-CALTTT-3}, brings out considerations on the role of unconventional properties like negative heat capacity on thermodynamics and statistical physics, and prompts the development of a novel minimal framework for thermodynamics that accounts for thermodynamic descriptions in between the purely statistical and phenomenological thermodynamics level. These results have a double outcome: (a) we develop a useful notion of thermodynamics beyond purely phenomenological thermodynamics with further possible applications; (b) we improve our understanding of equilibrium statistical mechanics, as we show how we can effectively apply equilibrium statistical mechanics to these unconventional systems if only we take a less stringent approach to features like equilibrium and stability. In addition, analysing SGS brings interesting considerations about the role of idealisations and what can we learn from idealised models via de-idealisation methods.

The structure of the paper is as follows. Section 2 introduces the background physics and unconventional behaviour of SGS. Section 3 introduces the debate around the applicability of statistical physics and thermodynamics in this exotic domain. Section 4 argues how equilibrium statistical mechanics can be applied to SGS to a certain scale and analyses the idealisations involved in SGS models. Section 5 fleshes out a minimal framework for thermodynamics that applies to SGS and argues that phenomenological thermodynamics does not work there.

\section{Atypical Features of Self-gravitating Systems}

Self-gravitating systems (SGS) are systems bound by their own gravity. Globular clusters are the prime examples of these systems. They are spheroidal conglomerations of self-gravitating stars, composed of tens of thousands to millions of elements. The physics of these systems can be in theory described to a high approximation via Newtonian gravitation. However, this million-body problem, as \cite{heggie_hut_2003} call it, is obviously practically intractable in that way. Because of this, statistical physics methods are employed to model these systems.\footnote{For a classic reference on galactic dynamics see \cite{binney2011galactic}.} SGS can be modelled like gases in which stars take the part of gas molecules, but this is not a perfect analogy. The crucial difference between SGS and conventional terrestrial systems to which statistical physics and thermodynamics are applied stems from the nature of the gravitational potential, which is long-range, and from the fact that gravity is universally attractive and does not possess any natural screening-off mechanism. 
As a consequence, unlike in the case of conventional gas molecules in which only short-range forces are dominant, the potential acting on any given constituent of SGS largely originates from the components that are not in its vicinity. These unconventional conditions determine several well-known kinds of problematic behaviour both from the point of view of statistical physics and thermodynamics:\footnote{\hl{In presenting these points I follow the exposition in} \cite{Callender2011-CALHAH} \hl{who introduced these features in the philosophical literature. These points can be found widely in the physics literature,} see \cite{lynden1977negative}, \cite{padmanabhan1990statistical}, \cite{binney2011galactic}, \citet[§13]{SwendsenExt}, \cite{heggie_hut_2003}. For an overview see \cite{Hutshortreview1997}.}

\begin{enumerate}
        \item \textbf{Divergences}: On the one hand, the gravitational potential has an infinite range, so the density of states and the (statistical mechanical) entropy diverge, and the microcanonical ensemble cannot be defined. On the other hand, short-range interactions are problematic too, because gravitational interaction is attractive without lower bound and the partition function in the canonical ensemble diverges.
        \item \textbf{Ensemble inequivalence}: The microcanonical and canonical ensembles are not equivalent within SGS, unlike in the case of conventional systems (at least in the thermodynamic limit).\footnote{We refer here to the thermodynamic limit $N, V \rightarrow \infty$ while $N/V$ is held constant.} Modelling SGS starting from different statistical ensembles leads to different descriptions and predictions. 
        \item \textbf{Non-extensivity:} Due to the long-range nature of gravity, functions such as energy and entropy are not extensive (in the statistical description), i.e. don't depend linearly on the size of the system, unlike for conventional systems. As reported by \cite{Callender2011-CALHAH}, many physics textbooks take extensivity as an axiomatic feature of statistical mechanics and thermodynamics. Even without taking it as axiomatic, it can be noticed that without extensivity we cannot derive what are regarded as key thermodynamic relations within statistical mechanics \hl{such as} $U = T S - P V + \mu N$.
        \item \textbf{Negative heat capacity}: SGS display negative heat capacity, that is $C_{V}=(\partial E/\partial T)|_{V}<0$, or $\dbar Q/dT<0$. It means that if the system gives out energy (e.g. the core of the cluster transfers energy to the halo) then its temperature increases. Positive heat capacity is often taken as a prerequisite of thermodynamics.
        \item \textbf{Instability and lack of equilibrium:} It is claimed that SGS are unstable and do not reach final thermodynamic/statistical equilibrium: it's entropically favourable for them to contract indefinitely. As it is conventional, `equilibrium' refers here to entropy extrema, i.e. vanishing of the first variation of entropy $\delta S$, while `stability' denotes the second variation of $S$ in the vicinity of equilibrium due to fluctuations.\footnote{See \cite{padmanabhan1990statistical} and \cite{heggie_hut_2003}. In a more fine-grained sense, (a) thermodynamic equilibrium is defined in terms of observables staying (roughly) constant, while (b) in statistical mechanics equilibrium can be either expressed in terms of the macrostate with the largest volume in the 6N-dimensional phase space (Boltzmannian approach), or in terms of the probability distribution being steady in time (Gibbsian approach).\label{equi}} A prime example of instability is the possibility of \textit{gravothermal catastrophe} (in the microcanonical ensemble): under the right conditions, due to negative heat capacity, the core of the system can get hotter when it gives out energy to the outer part. Then, if the energy absorbed by the outer part does not increase its temperature sufficiently to keep up with the increase in core temperature, the temperature gradient between the core and outer part steepens indefinitely and the core keeps contracting. An analogous phenomenon called \textit{isothermal collapse} happens in the canonical ensemble.\footnote{\cite{antonov1962solution} and \cite{lynden1968gravo} have been the first to study the topic.}
    \end{enumerate}

As we shall see in the next sections, the short-range divergence in point (1) and the ensemble inequivalence of point (2) are not deeply puzzling features and they can be addressed reasonably easily. The long-range divergence in point (1) constitutes a problematic idealisation, but we shall defer a proper review of it until §4.2. On the other hand, features (3), (4) and (5) are problematic features that substantively distinguish SGS from conventional systems and their presence calls for a reconsideration of the physics of these systems. 

The main questions to address at this point are the following: considering all these unconventional features, does thermodynamics apply to SGS in any way? What about statistical mechanics? Or should we merely accept that we can apply \textit{some} statistical methods to SGS without the whole underlying theory? The philosophical debate on the subject has mainly focused on these questions, which are a good starting point. We review it in the next section.

\section{The Callender-Robertson Debate}

In the philosophical literature the questions stated above have been discussed by \cite{Callender2011-CALHAH} and \cite{Robertson2019-ROBSAS-18}.\footnote{The topic is also covered in part in the papers by \cite{wallace2010gravity} and \cite{Callender2010PastHyp}.} These foundational issues are also often raised by physicists, although more sparsely.\footnote{The papers by Callender and Robertson are a useful reference resource in this respect.} This section summarises the debate and relates it to points (1)-(5). Before that, however, we draw a map of the domain of thermodynamics and statistical physics which will be useful to reconstruct the debate and further clarify the topic. 

We distinguish between three theories: (i) non-equilibrium statistical mechanics, a statistical theory about the dynamics of many-body systems out of equilibrium, represented for instance by kinetic theory and the use of Boltzmann equation; (ii) equilibrium statistical mechanics, applied to systems around statistical equilibrium where macroscopic quantities are roughly time-independent to a certain scale; (iii) phenomenological thermodynamics, a macroscopic theory formulated independently from microvariables and concerned with quantities like work and heat and with the operations we can perform with them.

Phenomenological thermodynamics is classical thermodynamics as specified by the phenomenological version of the laws of thermodynamics \citep{Wallace2015-WALTQC-2}. In a nutshell, the Zeroth Law defines the concepts of equilibrium and temperature in a broadly qualitative way, by claiming that if two systems are both in thermal equilibrium with a third system, then the two systems are in thermal equilibrium with each other. The First Law is about conserved energy transformations constituted by work and heat flow, i.e. $dU=\dbar Q + \dbar W$. The Second Law is understood via Clausius's or Kelvin's statements. If understood along these lines, we can think about thermodynamics as a broadly empirical and phenomenologically formulated theory, and possibly as a kind of control theory, as per Wallace: ``a theory of which transitions between states can be induced on a system (assumed to obey some known underlying dynamics) by means of operations from a fixed list" \citep[p. 699]{wallace2014thermodynamics}. 

Based on this distinction, we stress that the two sides of the following debate agree on the applicability of models and equations from non-equilibrium statistical mechanics while the disputed subject is rather the applicability of equilibrium statistical mechanics and (some concept of) thermodynamics.

\subsection{Robertson on self-gravitating systems}

\cite{Robertson2019-ROBSAS-18} argues that thermodynamics does not apply to SGS although (non-equilibrium) statistical mechanics does, based on the notion of thermodynamics she adopts, which is arguably phenomenological thermodynamics. She considers the puzzling behaviour of SGS and argues that, if we draw a clear distinction between thermodynamics and statistical mechanics, then the peculiar behaviour of SGS is not puzzling at all. In fact, rather than revising thermodynamics in light of the peculiar features of SGS, we should simply grant that (phenomenological) thermodynamics is not applicable to SGS while statistical physics can still be usefully and meaningfully applied in its non-equilibrium form. In taking this route of denying the applicability of thermodynamics to SGS she voices the scepticism expressed by physicists like \cite{binney2011galactic} on the use of thermodynamics in SGS, accounting at the same time for the fact that they still successfully apply statistical physics methods to these systems. Let's look at how Robertson characterises thermodynamics and statistical mechanics and her reasons for arguing that one theory applies while the other does not. 

Let's start with \textbf{thermodynamics}. First, she argues that ``thermodynamics is an abstract theory, that proceeds in ignorance of the constitution of the system, dealing instead only with macrovariables which obey the Four Laws" \cite[p. 1794]{Robertson2019-ROBSAS-18}, whereas statistical mechanics is based on the description of microvariables and their statistical and probabilistic distributions. She stresses that the physics of SGS does not really abstract away from these micro-level details (e.g. positions of the individual stars), and is instead based on equations such as the Boltzmann or Fokker-Planck equations. Hence we have a reason for rejecting the viability of thermodynamics in this area.

Second, she frames thermodynamics as essentially a theory about the behaviour of equilibrium states parameterized by macrovariables and about (thermodynamically quasi-stable) curves through the thermodynamic equilibrium state space. Positive heat capacity is also a crucial feature of this picture as it is crucial to securing stable equilibrium. This further motivates the claim that thermodynamics does not apply to SGS for the following reasons:

\begin{itemize}
        \item What would count as an equilibrium state in SGS? \cite{binney2011galactic} deny that there are any. Robertson points out that Dirac peaks and collapsed-core states could be theoretically proposed as equilibrium states for spherically symmetric SGS (as in Callender, \citeyear{Callender2011-CALHAH}), but these are (i) unphysical states,\footnote{It is easy to see how these are unphysical since the Dirac peak is a state in which all of the stars are placed in a single point, and the core-collapsed is not observed since we have instead gravothermal oscillations except for the cases in which a black hole forms, and in such cases Newtonian gravity breaks down.} and (ii) even if these were physically realistic states, they are just a single state and we need an entire state space for thermodynamics to work.
        \item Even if we could construct an equilibrium state space, SGS are unstable, because of the concavity of entropy and the negative heat capacity that leads to phenomena like the gravothermal catastrophe: small inhomogeneities are amplified rather than dissipated. Thus these systems do not return to equilibrium after a disturbance.
\end{itemize}

Third, if we regard thermodynamics as a control theory as previously defined, then thermodynamics is largely inapplicable to SGS since we cannot manipulate different parameters independently \cite[p. 20]{heggie_hut_2003}.

What about \textbf{statistical mechanics}? For the above reasons, Robertson argues that statistical mechanics does apply, but only as \textit{non-equilibrium} statistical mechanics. Most importantly, statistical mechanical equilibrium is never reached, but SGS still evolve \textit{towards} equilibrium, which leads to the claim that statistical physics applies in the non-equilibrium form. She suggests it is no surprise that this theory applies here because these systems are modelled as gasses in which the stars are gas molecules intrinsically identical and non-interacting, a claim supported by the idea that statistical physics essentially refers to the behaviour of the microvariables composing many-body systems.

Looking at the general picture, Robertson asks whether this situation is surprising. Can it be that statistical mechanics works while thermodynamics fails? Is not thermodynamic behaviour supposed to emerge from statistical mechanical behaviour? She argues that her conclusion is unsurprising, because the thermodynamic limit $N, V \rightarrow \infty$ (with constant $N/V$) does not exist for SGS, and it is in the thermodynamic limit that thermodynamic behaviour is supposed to emerge. Only in the limit statistical descriptions turn from probabilistic to categorical, certain statistical mechanical quantities become extensive thus coinciding with thermodynamic quantities, and the ensembles are equivalent. However, the conditions for the thermodynamic limit are not met in SGS. In particular, she singles out the following conditions, which do not apply: (i) interactions between distant particles must be negligible, (ii) interactions are stable.\footnote{More specifically, instability here refers to the facts that the gravitational potential is not bounded from below ($U \rightarrow - \infty$ as $r \rightarrow 0$) and that the gravitational potential violates the stability criterion that $U$ does not grow faster, as a function of $N$, than $N$.} Finally, it is \textit{equilibrium} statistical mechanics the theory from which thermodynamics is supposed to emerge.

\subsection{Callender on self-gravitating systems}

\cite{Callender2011-CALHAH} is more positive about the applicability of aspects of thermodynamics to SGS, although he concedes that the question remains open in several respects. The suggestion that thermodynamics is applicable to SGS does justice to the work of physicists such as \cite{lynden1968gravo}, \cite{chavanis2002thermodynamics}, and \cite{Katz2003-KATTOS}, who talk explicitly about the thermodynamics of SGS. Central to Callender's claim is the idea that key thermodynamic features such as thermodynamic equilibrium should be understood more liberally than is often done (cf. Callender, \citeyear{Callender2001-CALTTT-3}). His thesis that thermodynamics could apply to SGS if properly modified to address the strange behaviour displayed by SGS is grounded in his responses to the five unconventional features presented in §2. However, we stress that he is not specific on the meaning of thermodynamics he adopts, and if the project is the recovery of phenomenological thermodynamics or thermodynamics in another sense, so we remain neutral in this regard.

Concerning features (1) and (2), i.e. divergences and ensemble inequivalence, he deems them as unproblematic for the applicability of thermodynamics to SGS and hence dismissable.

On the one hand, the divergence determined by the nature of short-range gravitational interactions is fixed by introducing a short-distance cutoff in the models. The cutoff is needed to give meaning to the partition function as the gravitational interaction is attractive without lower bound and so a singularity problem arises in Newtonian gravitation. In galactic astrophysics, the introduction of the cutoff is easily justified given that the stellar radius provides a natural cutoff.\footnote{Furthermore, the introduction of the mean-field approximation helps to solve the problem.} It is also reasonable to say that this issue is not peculiar to these cases and thus not particularly concerning. On the other hand, Callender points out that the divergence introduced by the infinite (and unscreened) range of the gravitational potential is commonly solved in the physical literature on SGS models by putting the system in an ideal box.\footnote{As presented by \citet[p. 6]{chavanis2002gravitational}: ``We shall avoid the infinite mass problem by confining artificially the system within a spherical box of radius R. It is only under this simplifying assumption that a rigorous thermodynamics of self-gravitating systems can be carried out". The ideal box can be effectively replaced in a more realistic way by a potential well or tidal force \citep{yoon2011equilibrium}.} Concerning this last point we note that, unlike in the case of the short-range cutoff, it is reasonable to wonder whether this invoked idealisation is justified. We set aside this worry and return to it in §4.

Moving to ensemble inequivalence, recall that, unlike in conventional thermodynamics, the ensembles turn out to be generally inequivalent within SGS. This requires us to modify statistical mechanics within SGS, and, in general, the equivalence of ensembles can no longer be considered a general principle. Callender however suggests that this is not problematic and does not threaten the application of thermodynamics. \hl{He argues that} equivalence is a surprising feature of non-long-range interacting systems that emerges in the thermodynamic limit $N, V \rightarrow \infty$ (constant $N/V$), while there is no reason to think it is an essential prerequisite for statistical mechanics or thermodynamics.\footnote{A point we are sympathetic to.}

\hl{Concerning feature (3), i.e. non-extensivity, Callender points out that we can regain extensivity if we consider mean-field models and follow the so-called `Kac' prescription, appropriately rescaling the Hamiltonian for self-gravitating systems and applying the $N \rightarrow \infty$ limit (at fixed volume).}\footnote{The limit is applied to the Hamiltonian $H=\sum_{i=1}^{N} \frac{\boldsymbol{p}_{i}^{2}}{2m}+\frac{1}{2N}\sum_{i,j=1}^{N}V(\boldsymbol{q}_{i}-\boldsymbol{q}_{j})$. Note that only extensivity is regained whereas energy is still non-additive. See also \citet[p. 32]{campa2014physics} on the Kac prescription.} In that limit, energy and entropy become extensive (proportional to N) again.\footnote{In \S4.1 we present yet another sense in which the $N \rightarrow \infty$ limit matters to SGS physics.} An alternative limit is also introduced with a similar purpose by De Vega and Sanchez (\citeyear{devega2002statistical}, \citeyear{devega2006statistical}) who employ the dilute limit $N \rightarrow \infty$, $V \rightarrow \infty$, $N/V^{1/3}=fixed$ in astrophysics. An important point is that these limits are not the thermodynamic limit $N, V \rightarrow \infty$ (constant $N/V$) defined earlier. We mentioned how Robertson stressed the absence of this limit within SGS as one of the main reasons why the inapplicability of thermodynamics to SGS is unsurprising. Related to this point, Callender claims that the thermodynamic limit $N, V \rightarrow \infty$ (constant $N/V$) is neither necessary nor sufficient for thermodynamic behaviour.\footnote{See also \cite{butterfield2011less} on limits not being necessary for emergence.} Most importantly, he suggests that it might be wrong to think that one kind of limit is the right one generally speaking, and there is no \textit{a priori} reason to prefer that specific limit in every context. Instead, he suggests that we could take the limit $N \rightarrow \infty$ as the most suitable for SGS to recover thermodynamics since that limit allows us to regain extensivity. For this reason, we can call $N \rightarrow \infty$ the `gravitational thermodynamic limit'.\footnote{However, Callender is worried that this move could beg the question since we are deeming the $N \rightarrow \infty$ as the right thermodynamic limit in this context by starting from the assumption that extensivity must hold. We return to this point and refine this limit in §4.}

Concerning features (4) and (5), i.e. negative heat capacity and the strictly related instability and lack of equilibrium, the situation is more complex. Are there equilibrium states for SGS? Callender briefly considers the states mentioned before: a Dirac peak in which all the particles sit at the same material point when the (confined) system is attached to a heat bath, or an endlessly unbound shrinking-core system when the system is isolated. Should these exotic and arguably unphysical states count as equilibrium states in any way? Callender does not take a side and instead chooses another route. He suggests there is no equilibrium for SGS only if we adopt a strict notion of equilibrium. That is, he stresses that equilibrium is an idealisation and a matter of scales: ``Equilibrium, in thermodynamics and statistical mechanics, is an idealization. [...] there can't really be any \textit{truly} stationary states. [...] Equilibrium holds only with respect to certain observables, spatial scales and temporal scales" \cite[p. 967]{Callender2011-CALHAH}. Hence, considering the scales involved in SGS, he suggests taking the metastable states of the systems as a viable basis for equilibrium. 

However, granting that these could be regarded as equilibrium states under a more liberal notion of equilibrium, we can ask next whether these states are stable enough to allow for equilibrium in any meaningful way. And, indeed, later in the paper (§7.3) Callender himself raises the worry that the presence of negative specific heat could make equilibrium hopelessly compromised for systems like SGS, and he also questions the very postulation of negative specific heat: ``Are negative specific heats genuinely holding in equilibrium systems? Are they instead the result of idealisation? Or do they place peculiar constraints on how can combine thermodynamic macrostates?" \cite[p. 978]{Callender2011-CALHAH}. The negative heat capacity of SGS and the instability it entails is indeed the feature exploited by \cite{Robertson2019-ROBSAS-18} in arguing that, even if some sort of equilibrium could be found, SGS remain unstable and hence non-thermodynamic systems. 

Hence it seems that the physics of SGS could to some extent recover features like extensivity and quasi-equilibrium, but that any genuine thermodynamic description is \textit{prima facie} hindered by their inherent instability. The next sections pick up on the debate presented so far and advance it in several respects.

\section{Equilibrium Statistical Mechanics}

This section provides a first improvement of the debate so far. The applicability of key features like equilibrium and stability has been questioned in the context of SGS, hence the applicability of equilibrium statistical mechanics and thermodynamics has been challenged. Robertson argues that non-equilibrium statistical mechanics is the only working theory, whereas Callender voices optimism about the applicability of some kind of thermodynamics while leaving some legitimate open questions. 

This section advances our understanding of the physics of SGS by arguing in detail that \textit{equilibrium statistical mechanics} can be applied to SGS to a certain scale in a meaningful and useful way. §4.1 defends the scale relativity of equilibrium and argues for the applicability of equilibrium statistical mechanics in the context of SGS by showing how equilibrium and stability can fit into the physics of SGS in the form of metastable quasi-equilibrium states. Moreover, contrasting Robertson's claims that (a) SGS physics does not abstract away from the micro-details of the systems, and (b) the thermodynamic limit does not exist for SGS, it shows how macroscopic quantities can be defined within SGS and also how an appropriate limit for SGS can be found. §4.2 refines the arguments of the previous subsection. It shows how equilibrium statistical mechanics still applies to SGS even once we remove crucial idealisations involved in the models studied in §4.1 and move to more realistic ones. It does so by connecting in a novel way the physics literature on SGS with the philosophical literature on de-idealisations. The overall goal of §4 is to clarify which concepts and theories can be meaningfully applied in the atypical context of SGS, and to present how equilibrium statistical mechanics can be successfully used in situations in which we lack conventional equilibrium and negative heat capacity can be instantiated. This analysis also fosters a general reconsideration of concepts such as equilibrium and stability in statistical mechanics. Building on these results, §5 shows how part of the physics of SGS presented here can be considered as thermodynamic in character within a minimal notion of thermodynamics. The formulation of this framework and the consequent vindication of the notion of `gravitational thermodynamics' is the second main contribution of this paper.

\subsection{Statistical physics of idealised self-gravitating systems}

Let's start by stating our position in the context of the Callender-Robertson debate. We agree with part of Callender's conclusions, in particular we concur that both introducing short-distance cutoffs to avoid divergence of the partition function and the phenomenon of ensemble inequivalence are unproblematic steps. We also adopt for now the proposal of introducing an ideal box to model these systems to avoid density of state divergence, although the next subsection shows that things are more complex. We agree with the claims that extensivity can be regained in the appropriate but unconventional limit and that metastable states provide a useful basis for equilibrium within SGS, but we elaborate further on both topics. We also clarify the role of stability and negative heat capacity, showing that these systems are effectively stable and in equilibrium at the right temporal and spatial scale.

The absence of equilibrium and stability are central to Robertson's argument, hence these are our starting points. As maintained by Callender (\citeyear{Callender2001-CALTTT-3}, \citeyear{Callender2011-CALHAH}) equilibrium is generally scale relative, and perfect equilibrium is never reached even in conventional systems. Temporary fluctuations away from statistical equilibrium are to be expected and thus real systems are always in equilibrium relative to a certain scale. Drawing on this thesis, our central points are that (i) equilibrium states can be defined for SGS in the form of local equilibrium if considered in the appropriate regime and limit, as local entropy maxima instead of global maxima, and (ii) these equilibria are metastable states which are effectively stable relative to the right scale, as core collapse is triggered only in certain parameter regimes. While metastability refers to the stability of these states under minor perturbations, we call these entropy maxima `quasi-equilibrium states' to stress that they represent local equilibria that can nevertheless act as proper equilibria over the appropriate time scale, especially as SGS metastable states have extremely long lifetimes.

Let's look at the physics details to ground these claims. The focus is on isothermal spheres models of SGS.\footnote{\cite{antonov1962solution} proved that only spherically symmetrical states can potentially correspond to entropy maxima for these systems, hence only spherically symmetrical solutions are considered in general. The isothermal sphere is the simplest model of this kind. Cf. \citet[Appendix II]{lynden1968gravo} for the proof.} In the isothermal model the system is described statistical-mechanically by a distribution function $f=f_{0}exp(-2j^{2}E)$, with $f_{0}$ and $j$ as constants, and thus possess a Maxwellian distribution of velocities at any point. As shown in the seminal works by \cite{antonov1962solution} and \cite{lynden1968gravo}, and reported in \cite{chavanis2002thermodynamics}:

\begin{quote}
    Thermodynamical equilibrium of a self-gravitating system enclosed within a box exists only above a critical energy $E_{c}=-0.335GM^{2}/R$ or above a critical temperature $T_{c}=GMm/2.52kR$ and is at most a metastable state, i.e. a \textit{local} maximum of a relevant thermodynamical potential (the entropy in the microcanonical ensemble and the free energy in the canonical ensemble). For $T<T_{c}$ or $E<E_{c}$, the system is expected to collapse. This is called the ‘‘gravothermal catastrophe’’ or ‘‘Antonov instability’’ in the microcanonical ensemble and ‘‘isothermal collapse’’ in the canonical ensemble. \citep[§2.6]{chavanis2002thermodynamics}\footnote{\cite{antonov1962solution} showed that no globally maximum entropy state exists for particles of fixed total energy inside a box of finite $R$. Cf. \citet[Appendix III]{lynden1968gravo}.}
\end{quote}

Note that they call this feature `thermodynamic' equilibrium but what they are really employing is statistical mechanical equilibrium, as per the definitions in §2. Indeed the same equilibrium is elsewhere referred to as statistical equilibrium \cite[§4.3]{chavanis2006phase}, but most importantly this is the kind of equilibrium calculated e.g. from the phase volume and the distribution function.\footnote{Physicists working on the subject often call this thermodynamic equilibrium because they seem to identify thermodynamics with equilibrium statistical mechanics to a certain extent and because their focus is on studying thermodynamic potentials in equilibrium. Their rationale for labelling this `thermodynamic' is vindicated within our minimal approach to thermodynamics introduced in §5.}

The series of equilibria $(E, \beta)$ for classical isothermal spheres is represented in Figure 1, in which the (normalised) inverse temperature $\eta=\beta GM/R$ is plotted as a function of the (normalised) energy $\Lambda=-ER/GM^{2}$. Functions $\beta$ and $E$ are conjugate with respect to the entropy, $\beta=\partial S/\partial E$. The $\beta/E$ curve gives us information about the thermodynamic stability of isothermal spheres. Broadly speaking, it can be calculated that there is no stable equilibrium state above $\Lambda_{c}=0.335$ and $\eta_{c}=2.52$, but metastable states allowing for a scale-relative form of stable equilibrium can be found below those points. It is also worth stressing that the region of negative specific heats between CE and MCE is stable in the micro-canonical ensemble but unstable in the canonical ensemble \citep{padmanabhan1990statistical}, so we note that the relation between negative specific heat and instability is not straightforward and the existence of negative specific heat does not unambiguously block the existence of equilibrium (and on the other hand also positive specific heat can be defined for these systems).\footnote{More on negative heat capacity in §4.2.}

Figure 2 illustrates the behaviour of equilibrium states in the canonical ensemble. The metastable isothermal configurations are only local maxima since no global maximum can exist for these systems, and the value of energy or temperature is not sufficient to determine the evolution of the systems. A configuration with $\Lambda<\Lambda_{c}$ or $\eta<\eta_{c}$ can either reach metastable equilibrium or collapse depending on the position of the system with respect to the local entropy maximum. \cite{chavanis2002thermodynamics} studies the second variation of entropy and points out that isothermal spheres whose initial configuration is in $\triangle$ undergo collapse while the $\bullet$ configurations converge towards an equilibrium state.

\begin{figure}
    \centering
    \begin{minipage}{0.45\textwidth}
        \centering
        \includegraphics[width=1.1\textwidth]{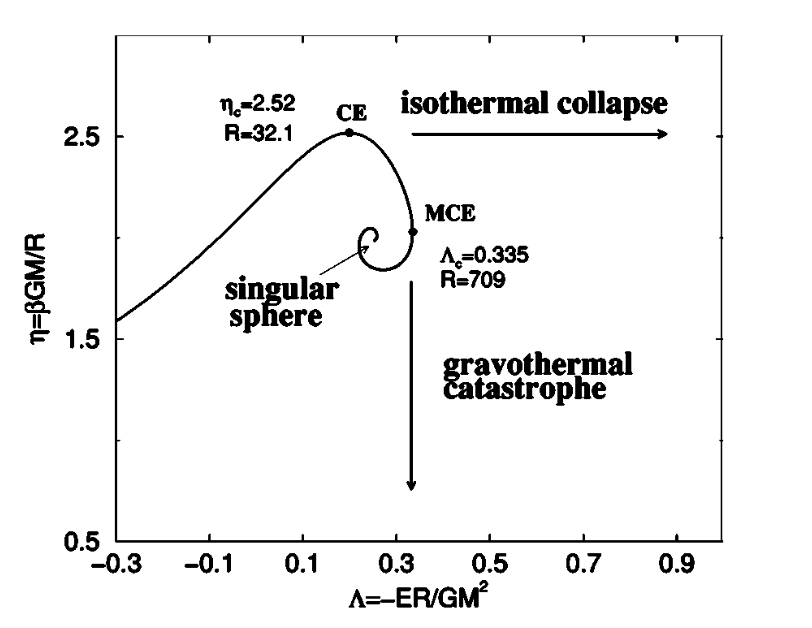} 
        \caption{\small Series of equilibria $(E, \beta)$ for classical isothermal spheres (a thermodynamic-kind diagram). (Chavanis et al., \citeyear{chavanis2002thermodynamics}, Figure 1)}
    \end{minipage}\hfill
    \begin{minipage}{0.45\textwidth}
        \centering
        \includegraphics[width=1.1\textwidth]{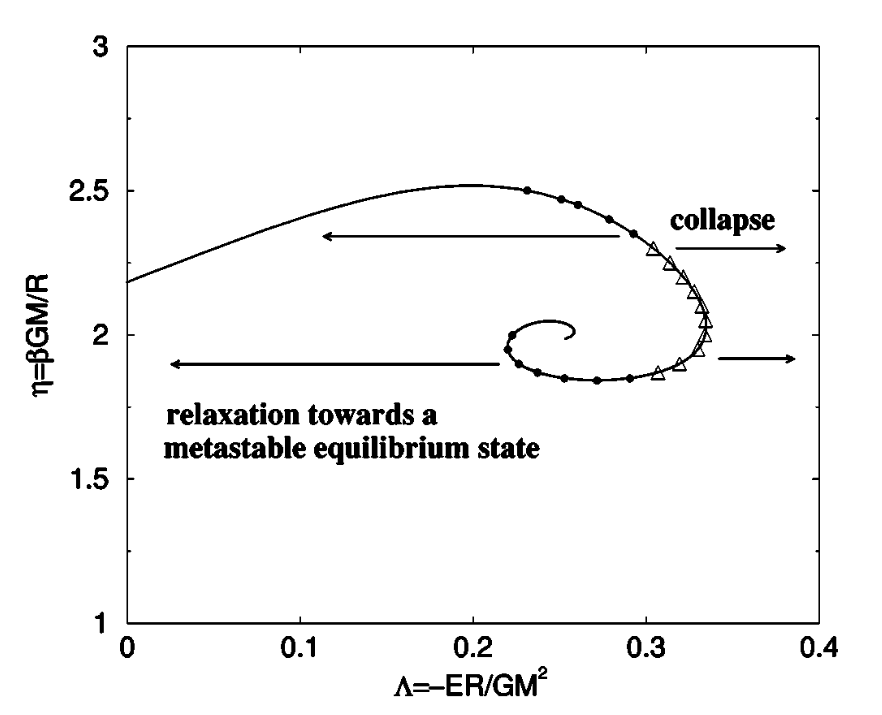} 
        \caption{\small Basin of attraction in canonical ensemble. Isothermal spheres after the first turning point of the spiral are unstable in this ensemble. Depending on their position on the spiral and the initial perturbation, they can either relax towards the local maximum of free energy with same temperature ($\bullet$) or undergo a gravitational collapse ($\triangle$). (Chavanis et al., \citeyear{chavanis2002thermodynamics}, Figure 16)}
    \end{minipage}
\end{figure}

This behaviour can be described at different levels of description, which also corresponds to different levels of precision and complexity. First, we can describe it within kinetic theory, where the series of equilibria can be calculated with some difficulty via the Fokker-Planck equation for the one-particle distribution function $f(\textbf{x},\textbf{p},t)$ (\citet[§4]{padmanabhan1990statistical}, \citet[§4.4, §7]{chavanis2006phase}):\footnote{See also \cite{chavanis2013kinetic}.}

\begin{equation}
    \dfrac{\partial f}{\partial t} + \textbf{v}\dfrac{\partial f}{\partial \textbf{x}} - \nabla \phi \dfrac{\partial f}{\partial \textbf{v}} = C(f),
\end{equation}
where $\textbf{v}=\textbf{p}/m$ and $\phi(\textbf{x},t)$ is the mean gravitational field produced by $f$,
\begin{equation}
    \phi(\textbf{x},t)=-G \int \frac{f(\textbf{y},\textbf{v},t)d^{3}\textbf{y}d^{3}\textbf{v}}{|\textbf{x}-\textbf{y}|}.
\end{equation}
Second, slightly more easily, we can calculate the variation of entropy within equilibrium statistical mechanics in the canonical and microcanonical ensemble, respectively from the density of states $g(E)$ and the partition function $Z(\beta)$ (\cite{horwitz1977steepest}, \cite{padmanabhan1990statistical}).\footnote{More precisely, for instance, \citet[§2]{Katz2003-KATTOS} employs a steepest descent technique starting from the density of states function. See also \cite{chavanis2002gravitational} and \cite{devega2002statistical}.} Finally, at an even simpler and coarser-grained level, we can maximize the thermodynamic potential \citep[§2-4]{chavanis2006phase} and employ Poincaré linear series of equilibria as developed by Katz (\citeyear{katz1978number}, \citeyear{katz1980stability}, \citeyear{Katz2003-KATTOS}).\footnote{See \citet[§17]{heggie_hut_2003} and \citet[§4.7]{chavanis2006phase} for an introduction.} This method then allows us to analyse the stability of SGS by focusing on the macro-quantities involved in our diagrams and on the topology of the curve $\beta/E$. Systems described by those parameters are stable along the curve until they reach a bifurcation or turning point, which is represented in the diagrams by vertical lines tangent to the equilibrium states line. We can use this method to study instabilities and predict in which parameter regimes they occur, matching the results of kinetic theory and equilibrium statistical mechanics based on $g(E)$ and $Z(\beta)$ but wholly relying on macro-level quantities defined over (scale-relative) equilibrium states within the theoretical framework of equilibrium statistical mechanics \citep[§7]{chavanis2006phase}.\footnote{
To anticipate, Katz's theory will provide in Section 5 the basis for the formulation of a minimal notion of thermodynamics that applies to SGS over merely statistical descriptions.}

This presentation shows that equilibrium and stability hold within SGS and that equilibrium statistical mechanics plays a crucial role in the physics of SGS. It is also important to stress the long lifetime of metastable states in SGS:\footnote{See also \citet[§2.6]{Katz2003-KATTOS}. \cite{chavanis2005lifetime} \hl{provides a method to estimate the lifetime of metastable states via Kramers formula. See} \cite{chavanis2024} \hl{for the detailed calculation of the lifetime of metastable states of a simple kind of SGS (a binary star) close to critical point}.}

\begin{quote}
    The lifetime of metastable states (local entropy maxima) scales as $exp(N)$ due to the long-range nature of the interaction. Therefore, the importance of these metastable states is considerable and they cannot be simply ignored. Metastable states are in fact stable [...]. \cite[pp. 135]{chavanis2005lifetime}
\end{quote}

The conclusion to draw is that metastable states provide stable equilibrium states in SGS within certain parameter regimes and time scales. These metastable states have long lifetimes compared to the evolution of these systems, a feature which can be related to the consideration that equilibrium is always relative to a certain time scale. Hence these technically quasi-equilibrium states can be regarded as effectively equilibrium states at the appropriate time scale. Collapses and instabilities take place and play an important role in the dynamics of SGS, but for the given reasons we can argue that equilibrium and stability can both be recovered within SGS in certain domains. It is worth noting that two distinct scales are operating in this context: a time scale (quasi-equilibrium) and a perturbation scale (metastability). These conclusions contrast Robertson's claims that equilibrium is not attainable within SGS and that only non-equilibrium statistical mechanics is employed and support instead the thesis that equilibrium statistical mechanics is applicable in this context, although this is not the full picture yet.

Having analysed equilibrium and stability, we now discuss two other key features related to the applicability of equilibrium statistical mechanics to SGS. In particular, we contrast the two interrelated Robertson's objections that (a) SGS physics does not abstract away from the micro-details of the systems, and that (b) the thermodynamic limit does not exist for SGS.

As reported in §3.1 Robertson argues that ``the thermal physics of SGS never abstracts away from to macroscopic bulk variables from the microvariables -- i.e. the position and momenta of the individual stars -- and probability distributions over these microvariables" \citep[p. 1795]{Robertson2019-ROBSAS-18}. In contrast, we highlight how macroscopic quantities actually play a role in the physical picture we have just presented, especially variables such as $E$, $N$, and $V$ that are used to define the density of states $\Omega(E, N, V)$ and entropy $S(E, N, V)$. These quantities are central to equilibrium statistical mechanics operations. 

\hl{Furthermore, other macroscopic quantities grounded in equilibrium statistical mechanics can be defined within SGS models. As mentioned,} \cite{Callender2011-CALHAH} \hl{stresses the ability of the mean-field approximation with $N \rightarrow +\infty$ at fixed volume (via the `Kac' prescription) to regain us extensivity of energy. But the use of such a limit in the mean-field brings more than that. As shown e.g. by} \cite{chavanis2006hamiltonian, chavanis2006phase}, \hl{by taking the $N \rightarrow +\infty$ limit (with fixed $\eta=\beta GM/R$ and $\Lambda=-ER/GM^{2}$) the mean-field approximation becomes exact, meaning that the total probability distribution for the system factorises, making the N-body distribution function a product of individual mean-field Maxwell-Boltzmann distribution functions. Hence we can apply an equilibrium statistical description, computing entropy maxima using the new distribution.}\footnote{\hl{On the use of `thermodynamic' functions for SGS in the $N \rightarrow +\infty$ limit see also} \citet[\S6]{tatekawa2005thermodynamics}, \citet[\S3.1.1 and \S3.2.1]{casetti2010solvable}, \cite{follana2000thermodynamics}.}

In this context, we can define several macroscopic quantities which are characteristic of the physics of equilibrium statistical mechanics, as Katz (\citeyear{Katz2003-KATTOS}, §3.1) remarks:  “Besides E, N, V and $S(E,N,V)$ there are other thermodynamic functions that give a physical content to the results [...] in particular the derivatives of S with respect to E, N, V” such as $\partial S/\partial E =\beta$, $\partial S/\partial N = -\alpha$, $\partial S/\partial V =\beta P_{b}$ (where $P_{b}$ is pressure on the boundary), from which follows:

\begin{equation}
    dS = \beta dE-\alpha dN + \beta P_{b}dV \quad \textrm{or} \quad \frac{1}{\beta}dS =dE-\frac{\alpha}{\beta} dN + P_{b}dV,
\end{equation}

from which we can define $T=1/\beta$ (temperature) and $\mu=\alpha/\beta$ (Gibbs chemical potential) which are global thermodynamic quantities of the whole system. Notice that the partial derivatives must be evaluated at the extremum, which highlights the role of equilibrium states. We take the definability of these quantities and relations as another proof of the applicability of equilibrium statistical mechanics over and above non-equilibrium statistical mechanics. Furthermore, this also supports the claim that $N \rightarrow \infty$ should be taken as the salient limit to recover macroscopic quantities within SGS instead of the $N, V \rightarrow \infty$ (constant $N/V$) thermodynamic limit. The latter might be the correct limit to adopt for short-range interacting systems if we want to recover macroscopic quantities, but there is no reason to believe it is true for every kind of system and thus we should instead be open to employing different limits in other contexts if we want to recover those kinds of coarser-grained features. This can help to address Callender's worry about the arbitrariness of adopting this limit.\footnote{We return to the justification of this limit in §5. In that section we also explain why the $N \rightarrow \infty$ limit deserves the name of \textit{thermodynamic} limit.}

Wrapping up, we have established several results. We have demonstrated that equilibrium, stability, extensivity, a viable `thermodynamic' limit, and certain macroscopic quantities and relations can all be recovered in SGS in certain regimes and provided certain idealisations. We thus made a strong case for the viability of equilibrium statistical mechanics in SGS, at least to an important extent, beyond non-equilibrium descriptions. However, two worries remain:

\begin{enumerate}
    \item The arguments provided rely on idealised models. In particular, real clusters are evidently not in boxes, but the box idealisation has been adopted from the outset without clear justification. Similarly, we have so far considered only collisionless systems. These features are drastic departures from realistic physical examples of SGS like globular clusters. Hence we ask whether they are justified, and what can we learn about actual systems from those descriptions.
    \item This section supported the application of equilibrium statistical mechanics to SGS. Hence it is reasonable to ask whether thermodynamics still has a place in the physics of SGS, especially since we countered Robertson's main reasons against the applicability of thermodynamics.
\end{enumerate}

§4.2 addresses the idealisations, §5 responds to the second group of questions and proposes a framework that can cash out SGS thermodynamics.

\subsection{Removing idealisations}

The model in §4.1 relied on the strong assumption that the systems are confined in a box of fixed volume and a given radius. However, astrophysical systems are evidently not confined by boxes and there are no real walls against which the stars can bounce. Furthermore, it is not clear whether there is any kind of strong external potential that can approximately play the same role and effectively confine these systems like a box. Hence it is reasonable to question whether this idealisation is justified. For example, Touma and Tremaine complain that: 

\begin{quote}
    To make progress, the usual approach is to introduce artificial cutoffs by confining the N-body system in a spherical box [...] Although instructive and elegant, these models leave one with the nagging question of what all of this has to do with actual self-gravitating systems in the real world. What remains at the end of the day are robust results on the thermodynamics of artificially imprisoned and mutilated self-gravitating systems, more tentative and largely numerical results on the evolution of realistic systems, and heuristic rules relating the properties of the former to the latter. \cite[p. 2]{touma2014statistical}
\end{quote}

The box idealisation is important because it ensures that the model displays meta-stable equilibria as we state in more detail below. Moreover, in §4.1 we did not consider the possible impact of collisions, and we may ask whether factoring collisions in would hinder stability and equilibrium. If these idealisations are not justified then it is not clear whether we are learning anything useful about real systems. To tackle the issue we provide a clear criterion for testing the justifiedness of an idealisation in the form of Earman's principle (as \cite{jones2006ineliminable} and \cite{landsman2013spontaneous} call it) and apply it to our case study. The principle provides a necessary (but not sufficient) condition for successful de-idealisations that is particularly useful for our analysis, hence we use it as a benchmark.

\begin{quote}
    \textbf{Earman's principle:} ``No effect [predicated by a model] can be counted as a genuine physical effect if it disappears when the idealizations [of the model] are removed".\footnote{As quoted in \citet[p. 5]{fletcher2020principle}.} \citep[p. 191]{earman2004curie}
\end{quote}

In particular, since the application of equilibrium statistical mechanics advocated in §4.1 was based on showing that equilibrium and stability hold for SGS in the right regime, in what follows we test whether equilibrium and stability disappear once we remove the box idealisation and we factor in the effects of collisions on the evolution of SGS. We first present qualitatively the effects of these idealisations on SGS models under different conditions to show how they influence the case study and then analyse what happens when we remove the idealisations. We argue that metastable quasi-equilibrium states do not disappear after relevant de-idealisation, rather they are still present in the more realistic King models and are also displayed by the core of real globular clusters, which approximate the isothermal models of §4.1. Hence, by Earman's principle, these observations support the validity of the conclusions of the previous section. Additionally, this analysis provides a more refined account of how equilibrium statistical mechanics applies to SGS.

We start by highlighting that the box idealization has clear effects on SGS models under certain conditions. If we model SGS as systems confined in a box that allows the stars to bounce against the walls without losing energy, then we can always make the system stable and in equilibrium by restricting the radius of the box enough \cite[p. 177]{heggie_hut_2003}. If the radius is small enough, kinetic energy dominates and the effect of gravity is negligible, thus the system behaves like an ideal gas and is not in virial equilibrium (and the heat capacity is positive). As \cite{heggie_hut_2003} remark, systems in these models are stable, as proved by \cite{lynden1968gravo}. Thus, if the box is small enough the behaviour of the system is heavily influenced by it and equilibrium statistical mechanics `trivially' applies. What about bigger boxes?\footnote{For a more quantitative description of the scenario see \citet[§8, §17]{heggie_hut_2003} and \citet[§4.4]{padmanabhan1990statistical}.} If the box is large enough then the stability depends less trivially on the energy of the system, as per Figures 1 and 2 above. If the energy is high enough, then as before the kinetic energy dominates over the gravitational effects and we have metastable quasi-equilibrium states as argued in the previous subsection. However, the idealisation is still influencing the model's behaviour and thus it is reasonable to inquire about its justifiedness. Finally, in the isothermal models of §4.1 there was no mention of collisions and interactions between stars, and collisionless evolution naturally renders stability easier to secure. It is reasonable to ask whether taking into account interactions makes stability unattainable.

Let's then follow Earman's principle and analyse what happens once we de-idealise. The first step we take in the removal of the box idealisation is to employ the King model instead of the isothermal sphere model confined in a box \citep{king1966structure}. We adopt this model because it can be regarded as the simplest kind of realistic model of a globular cluster that is not artificially confined and does not have pathological behaviour like the infinite mass of the basic isothermal model \citep{elson1987dynamical}. Adopting this model is thus the easiest way to de-idealise the model studied previously to apply Earman's principle. 

In a nutshell, the King model is a spherically symmetric modified isothermal model. In this model, stars are not mechanically confined by any boundaries and are allowed to escape (although the system is still generally bound gravitationally). However, the system is still parametrised by a radius to avoid the problem of infinite mass and the divergence of the density of states: if stars that evaporate are still considered in the system, then the system's extension will tend to infinity and the density of states will diverge making the statistical mechanical description not well-defined. To make the system statistically tractable, the model introduces a cutoff radius which is determined by the escape energy of the high-energy stars that evaporate (\cite{meylan1997internal}, \cite{chavanis2006phase}). The distribution function of the King model is:

\begin{equation}
  f =
    \begin{cases}
      f_{0}(exp(-2j^{2}E)-exp(-2j^{2}E_{0}) & \text{if $E<E_{0}$},\\
      0 & \text{if $E>E_{0}$},
    \end{cases}       
\end{equation}
where $E_{0}$ is the escape energy. The model has a cutoff radius and a core or King radius $r_{0}$ defined by $r_{0}\equiv \sqrt{9\sigma^{2}/4\pi G \rho_{0}}$ where $\rho_{0}$ is the central density and $\sigma$ is the dispersion. $r_{0}$ is ``the radius at which the projected density of the isothermal sphere falls to roughly half of its central value" \cite[p. 305]{binney2011galactic}. We thus have a cutoff radius to allow us to model the system statistical-mechanically but we don't have a confining box anymore. 

King models are important and widely used in the physics of SGS as they are well-confirmed experimentally and thus realistically model globular clusters \citep{elson1987dynamical}. King models also approximate more recent and developed `lowered isothermal' models that fit observational data even better (\cite{gieles2015family}, \cite{zocchi2016testing}), but are more tractable.

What happens to these systems now that the box is removed? We focus on globular clusters in the King model and present their evolution taking also into account the effects of collisions and interactions.\footnote{\hl{In the context of globular clusters, by collision we mean close gravitational encounters.}} These models behave like real globular clusters, and these astrophysical objects are believed to naturally tend towards core collapse in normal conditions once encounters take place. But let's first distinguish two time scales, the crossing time $t_{cross}= R/v$ relating size of the system and velocity of stars and the relaxation time $t_{relax}\sim \frac{0.1N}{lnN}t_{cross}$, where $N$ is the number of stars in the system. The crossing time is the time needed for a typical star to cross the galaxy once and is ``the shortest time scale on which the system as a whole can react to global changes in its potential", the relaxation time is ``the time scale in which the cumulative effect of two-body encounters can alter the individual stellar orbits significantly" \citep[p. 566]{elson1987dynamical}. The relaxation time is significant for globular clusters and on this time scale encounters drive the evolution of stellar systems by different mechanisms (\cite{meylan1997internal}, \citet[§7]{binney2011galactic}). 

The evolution of globular clusters is mainly driven by two-body relaxation, a type of gravitational encounter. These encounters, together with other processes, drive the system towards states with denser cores and low-density halos, which also have higher entropy. Some of the other mechanisms involved are the following. First, due to the equipartition of energy (and the virial theorem), massive stars with less kinetic energy fall deeper into the gravitational potential well while less massive stars diffuse towards the outer parts, bolstering the core-halo structure.\footnote{However, note that mass segregation is not \textit{necessary} for core collapse, as collapse can take place even in systems composed of stars of equal mass \citep{heggie_hut_2003}.} Second, encounters lead to evaporation, that is the stars with higher velocity can escape the cluster, as mentioned earlier, decreasing the total energy. Third, within realistic globular clusters, the tidal field of the galaxy also has a role in evaporation, as it strips away stars from the clusters. Relatedly, star escape is also enhanced by tidal shocking.\footnote{On the effects of evaporation see in particular \cite{king1966structure}, \cite{chernoff}, \citet[§2]{elson1987dynamical}, and \citet[§5]{spitzerdynamical}.} The effect of evaporation is particularly relevant for clusters with concentration parameter $c<2.1$ \citep{Inagaki_1988}. The concentration parameter characterises the profile of the King model and is $c = log (r_{t}/r_{c})$, where $r_{t}$ is the truncation radius and $r_{c}$ is the core radius.

Hence the system naturally slowly evolves towards a core-halo structure with a progressively shrinking core, and overall there is a decrease in the radius of the cluster and an increase in the concentration parameter $c$. For $c<2.1$ the cluster follows the King sequence closely, whereas for larger values of $c$ the King models become unstable.\footnote{As Elson et al. remark: “This is in reasonable agreement with observations of most globular clusters, as well as with numerical calculations of globular cluster evolution in the Fokker-Planck approximation, which show that an evolving cluster follows the King sequence closely as long as $c<2.1$” \cite[pp. 568-9]{elson1987dynamical}.} This instability, powered by the negative heat capacity, sets the final stage of the core collapse in the form of gravothermal (or isothermal) collapse \cite[§4.9-4.10]{chavanis2006phase}.\footnote{See also \citet[p. 5]{chavanis2015darkmatter} for a description of self-gravitating systems within the King model (i.e. without the box). It should be noted that the same phenomenon happens in the box-confined idealised case when the box is big enough that kinetic energy does not dominate and thus it becomes entropically favourable to form a core-halo structure.} Following this runaway process, the system collapses up to a point in which binary stars' interactions emit enough energy to make the system re-expand and this triggers gravothermal oscillations.\footnote{See \citet[Part VIII]{heggie_hut_2003} on post-collapse evolution. Two classic works on the topic are \cite{cohn1984there} and \cite{hut1983binaries}.}

Does this prove that \textit{realistic} SGS such as globular clusters are essentially always unstable and not in equilibrium as they constantly evolve towards core collapse? The answer is no. These effects take place relatively slowly and on the relevant time scale these systems progress through a series of metastable quasi-equilibrium states. For instance, as \citet[p. 33]{chavanis2002gravitational} remarks: ``evaporation is a slow process so that a globular cluster passes by a succession of quasi-equilibrium states corresponding to truncated isothermals".\footnote{See also \citet[p. 255]{Katz2003-KATTOS} who makes the same point within King models.} \cite{elson1987dynamical} raise a similar point. They first claim that stars in the high-velocity tail of the clusters are continuously torn out by the tidal field of the galaxy, causing the shrinking of the radius and the increase in $c$, but then they point out that “Although a cluster never attains a final equilibrium state, its evolution can be represented as a series of near-equilibrium configurations described by King models” \citep[pp. 568]{elson1987dynamical}, and stress that this is in agreement with observations and numerical calculations as long as $c<2.1$ (i.e. the stable regime). Thus, while it is true that these systems naturally progress towards core collapse and that they are unstable in certain regimes, this does not rule out the existence of effectively stable equilibria at the appropriate scale.  

Hence we can use equilibrium models for relaxed clusters over the relevant time scale, and these models are useful tools in the physics of SGS. \cite{katz1978steepest} build a whole statistical mechanics for these types of models while \cite{katz1980stability} provides a coarse-grained stability analysis for King (and other) models based on $(E, \beta)$ diagrams and the Poincaré linear series of equilibria method presented in the previous subsection (Fig. 3). Therefore we can argue that features like equilibrium and the use of equilibrium statistical mechanical methods based on macro-quantities are not mere artefacts of the idealised isothermal box-confined model described earlier, but rather are available in more realistic models after de-idealisations. In the case of the King models, the system can evolve through the diagram through a series of quasi-equilibria as it loses energy until it reaches a turning point, which is represented in our diagrams by vertical lines tangent to the equilibrium states line. At that point, it loses stability.

\begin{figure}
    \centering
    \includegraphics[width=0.7\linewidth]{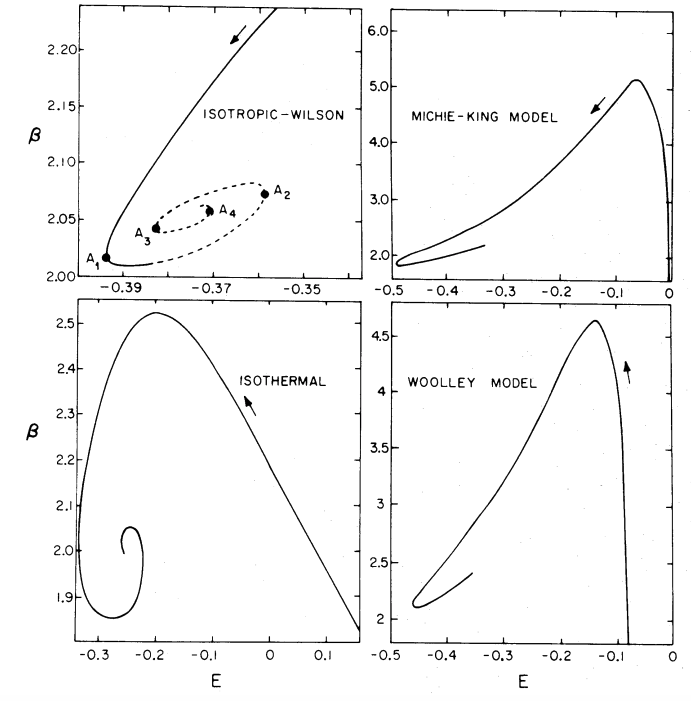}
    \caption{Series of equilibria $(E, \beta)$ for different models. (Katz, \citeyear{katz1980stability}, Figure 1)}
\end{figure}

Before concluding, two other points are worth mentioning to make the argument of this subsection even more robust. First, the analysis by \cite{katz1978steepest}, which starts from the premise that models like the King one are definitely more realistic than artificially box-confined isothermal models, eventually shows that ``As far as the core of the cluster is concerned, there is little difference between both models in terms of the stability conditions." \citep[p. 309]{katz1978steepest}. That is, the very analysis of §4.1 remains valid for the more realistic King models if we focus on the core. However, this observation is not surprising and relates to our second point. That is, both King models and even more developed and realistic models like the `lowered isothermal' models by \cite{gieles2015family} have isothermal cores.\footnote{See \citet[§22.3]{bertin2014dynamics} on the approximate isothermality of the core of Kings models.} Hence, at the right parameter regime, the core of these clusters approximates the isothermal profile discussed in §4.1 and we can study the core of these clusters as an isothermal system slowly moving through quasi-equilibrium metastable states.\footnote{See \citet[pp. 169-170]{heggie_hut_2003} on the interaction between the core and the outer part of the system in this situation.}

To sum up, we have provided several reasons in support of the robustness of the analysis carried out in the previous subsection. We have done so by following Earman's principle and showing how equilibrium and stability survive once we move from idealised to more realistic models through de-idealisations. In particular, we removed the box idealisation and the collisionless assumption. We demonstrated that King models, which are some of the simplest examples of non-artificially-confined models, still display quasi-equilibrium metastable states and can be modelled using the kind of coarse-grained methods used in §4.1 even when encounter effects are factored in. Then we also pointed out that the cores of realistic models of globular clusters approximate the isothermal sphere profile and therefore this supported in an even stronger way the validity of the analysis of §4.1, since the isothermal model employed in that context remains a valid description of a part of real clusters. Therefore the applicability of equilibrium statistical mechanics to SGS is not an artefact produced by the unrealistic nature of the model we employed in §4.1, but rather we are justified to believe that this theory plays an actual role in the physics of SGS.\footnote{Another realistic case in which the equilibrium statistical description and the series of equilibria is valid is studied in \cite{bertin2003thermodynamic} for partially relaxed stellar systems.}

More generally, this subsection has also developed a more refined description of how equilibrium statistical mechanical behaviour is instantiated by realistic SGS and has provided a more accurate description of the scales and regimes at which this analysis is valid within globular clusters.

Before moving on, a note on the role of negative heat capacity within the statistical physics of SGS is in order. As pointed out in §3.2, \cite{Callender2011-CALHAH} asks whether negative specific heat would ultimately be unavoidably problematic for equilibrium and whether negative specific heat can genuinely hold in equilibrium systems. This section dispelled these worries. Core collapse and negative heat capacity are genuinely displayed by SGS, but only in certain regimes, and only under certain conditions negative heat capacity makes the system completely unstable. Hence SGS can still be described as evolving through a series of effective equilibria in the right regime even if negative heat capacity can be instantiated at certain points.

\section{Minimal Thermodynamics}

The previous sections have supported the applicability of the core scaffolding of equilibrium statistical mechanics to SGS. However, equilibrium statistical mechanics is compatible with thermodynamics and indeed thermodynamics is usually taken to emerge from it. And, in the last section, we have been able to define within the equilibrium statistical mechanics of SGS quantities, relations and a limit that are usually regarded as `thermodynamic'. Here we return to the central question of the whole debate and ask whether thermodynamics does apply to SGS and, if yes, in what sense. \hl{Based on the results of the previous sections, we argue that a \textit{`minimal' thermodynamic level of description} is available in the case of SGS, even if the phenomenological thermodynamics level of description in terms of e.g. heat, work, and Clausius' second law of thermodynamics expressed via heat transfer has no practical use here.

In a nutshell, we argue that we can devise macroscopic descriptions and models of SGS as well as explanations of their behaviour via macroscopic quantities like state functions in terms of (e.g.) energy and temperature over and above the purely statistical descriptions in terms of (e.g.) partition functions. We call them minimal thermodynamics descriptions and explanations: ‘thermodynamics’ because they involve macroscopic descriptions of these systems above the level of description of the partition function and density of states, ‘minimal’ because this is still not the level of description of phenomenological thermodynamics.  

This middle-way level of description abstracts away from the statistical descriptions of SGS by introducing new macroscopic quantities allowing for simpler novel explanations of the behaviour of SGS, largely autonomous from the lower-level more complex statistical level. This provides a new perspicuous conceptual framing of the physics of SGS and a finer-grained account of all the levels at which it operates, and vindicates a notion of gravitational thermodynamics. There are four levels of descriptions within statistical and thermal physics, corresponding to phenomenological thermodynamics, minimal thermodynamics, equilibrium statistical mechanics, and non-equilibrium statistical mechanics. Based on \S4 we say that only the last three apply to SGS. These are distinct levels and we can provide different types of explanations of the behaviour of SGS at each level. Merging together two or more of these levels would render our understanding of the physics of SGS less fine-grained.}

\subsection{\hl{Introducing minimal thermodynamics}}

 \hl{We introduce the notion of minimal thermodynamics in general terms. The next subsections clarify the framework by applying it to SGS. Minimal thermodynamics refers to a descriptive framework situated between statistical mechanics, which uses ensembles, and phenomenological thermodynamics, which focuses e.g. on heat exchange. We frame minimal thermodynamics not as an additional theory but rather as an additional level of description that can be embedded in the theoretical framework of equilibrium statistical mechanics.}\footnote{\hl{Although we are not committed to a strict distinction between theories and levels of description, an intuitive way to see how they differ is by noticing how we can have different levels of description within the framework of one theory. For example, the same system in the same theory can be described at different levels of abstraction} (Franklin and Knox, \citeyear{Franklin2018-FRAEWL}).} \hl{We include it there and not within phenomenological thermodynamics because we understand the latter as a `control' theory focused on operations we can do on and between systems involving concepts like heat exchange, work, or Carnot cycles.}\footnote{\hl{However, this is mostly a semantic point and our presentation does not hinge on that: in principle, one would be free to follow} \cite{Callender2011-CALHAH} \hl{and relax our concept of phenomenological thermodynamics and argue that minimal thermodynamics is another lower level of description within the theory of phenomenological thermodynamics in addition to the traditional level of phenomenological thermodynamics just described. Hence we stress that our approach is consistent with Callender's proposal of relaxing our notion of thermodynamics.}}

The quantities of the minimal thermodynamic description are coarse-grained quantities like energy, temperature and entropy that can be derived within equilibrium statistical mechanics but can enter higher-level explanations of the behaviour of SGS above the mere lower-level statistical mechanical behaviour expressed in terms of micro-variables and distribution functions. At the same time, minimal thermodynamics is not at the same level of description as phenomenological thermodynamic descriptions, which are even more abstract. Hence `minimal thermodynamics' captures a distinctive level of description and layer of explanations. We stress that this notion is meant to be general and not restricted to the case of SGS. Here is the general framework (see also Table 1): 

        \begin{enumerate}[I.]
            \item \textbf{Quantities:} Minimal thermodynamics abstracts away details about the composition of the system, in contrast to the description of statistical mechanics expressed in terms of dynamics of micro-variables. It is based on macroscopic quantities that are state functions defined within equilibrium statistical mechanics ($E$, $T$, $V$, ...), often in the limit. Unlike phenomenological thermodynamics, it is not based on empirically formulated descriptions in terms of $\delta Q$ and $\delta W$ involved in transformations on systems.
            \item \textbf{Explanations:} Within minimal thermodynamics we can deliver `thermodynamic' explanations based on the behaviour of these macroscopic quantities. They are formulated within the theoretical framework of equilibrium statistical mechanics but are higher-level non-statistical explanations.
            \item \textbf{Autonomy \hl{and novelty}:} Minimal thermodynamics is partly autonomous from the statistical lower-level description, although it is based on quantities derived from it. Minimal thermodynamics concerns the evolution of the macroscopic quantities irrespective of their statistical underpinning. \hl{Minimal thermodynamic explanations can be novel insofar as the explanations we devise via the macroscopic variables above offer additional explanations of a given phenomenon which were not available in the same form at a lower level of description.}
            \item \textbf{Trade-off:} Minimal thermodynamics explanations and predictions are simpler but more limited than the statistical mechanical explanations and predictions of the same phenomenon.
            \end{enumerate}

\hl{The features of this framework align with the widespread trend in the philosophy of science that emphasizes the crucial role of abstractions in scientific explanations} (\cite{Batterman2002-BATTDI, Batterman2009-BATIAM-4, Batterman2021-BATAMW}, \cite{Strevens2008-STRDAA}, \cite{weslake2010explanatory}, \cite{Morrison15}). \hl{The general idea is that more abstract and less precise explanations can be nonetheless more powerful than more detailed explanations in certain respects, for instance in virtue of being more practically useful. The framework of minimal thermodynamics is especially consistent with the account of abstraction developed by} \cite{Knox2016-KNOAAI-3, knox23} and \cite{Franklin2018-FRAEWL}. \hl{They defend the usefulness of abstractions and in particular abstractions obtained by change of variables that allow us to screen off certain details. Such abstractions change the level of description at which a certain system is described, and enable a simpler representation of the system, facilitating novel and more effective explanations of the same phenomena. Their examples include changes of variables in the descriptions of harmonic oscillators, which lead to simpler equations, or—more interestingly—changes of variables that make the descriptions of phonons more perspicuous. In the case of SGS, we argue that introducing diagrams like those in Figures 1--3, where SGS are modelled solely in terms of energy and temperature, allows us to analyse certain aspects of their behaviour using the simple series of equilibria method. This approach provides simpler and useful models that are not available at lower, less abstract levels of description. This is the sense in which focusing levels of description is particularly interesting in the present context.}

\renewcommand{\arraystretch}{1.4} 

\begin{table}
\caption{Theories and levels of description.}
\label{demo-table}
\begin{tabular}{ | >{\raggedright\arraybackslash}m{3cm} | >{\raggedright\arraybackslash}m{5cm} | >{\raggedright\arraybackslash}m{2.7cm} | } 
 \hline
 \textbf{Theory} & \textbf{Level of Description} & \textbf{Variables} \\ \hline
 Phenomenological Thermodynamics & Empirical descriptions in terms of macroscopic transformations on systems & $\delta Q$, $\delta W$, $...$ \\ \hline
 \multirow{2}{*}{Equilibr Stat Mech} & \textit{Minimal Thermodynamics:} Coarse-grained state functions and relations, etc. & $E$, $T$, $V$, $P$, $S$, $...$ \\ \cline{2-3}
  & \textit{Equilibrium Stat Mech:} Canonical and microcanonical ensemble, density of state, partition function, etc.  & $g(E)$, $Z(\beta)$, $...$ \\ \hline
 Non-Equilibrium Stat Mech & Evolution of one-particle distribution functions via Boltzmann equation, etc. & $f(\textbf{x},\textbf{p},t)$, $...$ \\
 \hline
\end{tabular}
\end{table}

\subsection{\hl{Minimal thermodynamics of self-gravitating systems}}

We now show how the minimal framework for thermodynamics just introduced can account for certain characteristic aspects of the physics of SGS. Applying minimal thermodynamics to our case study is also the best way to spell out the general description of the framework just presented.

A specific and clear example of the application of minimal thermodynamics is the descriptions and explanations of SGS based on the diagrams in Figures 1, 2 and 3. Katz (\citeyear{katz1978number}, \citeyear{katz1980stability}, \citeyear{Katz2003-KATTOS}) developed Poincaré's linear series of equilibria theory to analyse the stability of SGS in terms of quantities such as $E$ and $\beta$.\footnote{Following the approach by \cite{antonov1962solution} and \cite{lynden1968gravo}.} Basically, turning points in the $(E, \beta)$ phase diagram denotes instabilities:
    \begin{quote}
         The stability or the number of unstable modes can be deduced from the topological properties of series of equilibria, i.e., \textit{from purely thermodynamic considerations} \citep[p. 241, italics added]{Katz2003-KATTOS}
    \end{quote}
    
    Thus, looking at macro-level quantities provides a platform for developing an explanation of the stability of SGS and their evolution into a core-halo structure from a distinctive higher-level perspective. As stressed by \citet[§4.4]{chavanis2006phase} and \citet[§2.4, §3.3]{Katz2003-KATTOS} the results we obtain with these methods are compatible with the results obtained via kinetic theory and numerical methods and thus are in principle derivable also from that more detailed lower-level perspective. Hence we have at least two distinct dimensions of descriptions and explanations working at the same time, one characteristically statistical mechanical and the other thermodynamic, in the sense of being coarse-grained with respect to micro-variables and e.g. distribution functions.\footnote{More specifically, as mentioned in §4.1, the statistical-based dimension of explanation comprises both non-equilibrium and equilibrium statistical explanations.} This shows how Katz's approach adheres to points (I)-(II) of minimal thermodynamics. 
    
    Concerning (III), we point out that these higher-level descriptions and explanations are partly autonomous from the statistical level: they are based on a diagram obtained by maximizing statistical entropy and so the $(E, \beta)$ description is derived from the statistical mechanical description, but Katz's explanation of instability is purely expressed in terms of the behaviour of these macro-quantities independently from their underpinning. \hl{Relatedly, these explanations can be novel in the sense of \S5.1 for the reasons given in the previous paragraph: introducing the minimal thermodynamic description in terms of state variables allows for building new models of SGS and a novel dimension of explanations for SGS behaviour on top of the kinetic theory level and the ensemble level.} Relating this point to the literature on inter-level relations in science, this type of independence from the micro-level and novelty could be framed as emergent behaviour, in a sense of emergence compatible with reduction, as these descriptions are in principle derivable.\footnote{See for instance \cite{butterfield2011less} and \cite{Franklin2018-FRAEWL}. More precisely, it could be classified as `coarse-grained emergence', following \citet[p. 39]{Palacios2022-PALEAR}.}
    
Furthermore and relatedly, concerning point (IV), as Katz remarks the higher-level explanations are more limited: 

     \begin{quote}
        The thermodynamic criterion has a number of limitations. One rarely calculates all the sequences of equilibrium and therefore some bifurcations may not show up because the branch points are missing. Thus equilibria might become unstable and the system might choose to be in a more stable state which has not been calculated. […] Another general limitation of any thermodynamic criterion of stability is that we learn little about the nature of instabilities, triggering mechanisms, and what becomes of stable states which evolve through a series of quasi-equilibria along the linear series up to and beyond the limit of instability. \cite[p. 245]{Katz2003-KATTOS}
    \end{quote}

However, they are simpler to employ as it is easier to explain the behaviour and shape of globular clusters by running a stability analysis solely based on the $(E, \beta)$ curve than by solving the Fokker-Planck equation. As Yoon et al. stress, ``the thermodynamical approach to describe the self-gravitating system [...] is useful to understand important physics using much less expensive computational resources than the numerical simulation" \cite[p. 2737]{yoon2011equilibrium}. So there is a trade-off between higher practical utility and less detailed explanations. Also, the trade-off between more abstract and more detailed descriptions arguably matches the one we normally find in confronting thermodynamic and statistical mechanical explanations of the same phenomena, hence this further vindicates the thermodynamic character of minimal thermodynamics.

The approach just discussed is a specific example, but is representative of the general fact that in certain regimes SGS can be expressed in terms of macro-level quantities as presented in the previous section. For instance, virial equilibrium (as $2K + U = 0$), negative heat capacity (as $C_{V}=(\partial E/\partial T)|_{V}<0$), and temperature ($T^{-1}(E)=\beta(E)=\partial S(E)/\partial E$) are central to \cite{lynden1968gravo} and the related literature. Also, as presented in \S4.1, `thermodynamic' functions in the sense of minimal thermodynamics are used by e.g. \cite{tatekawa2005thermodynamics} and \cite{casetti2010solvable} within a variety of SGS models. Furthermore, relations like $dE = TdS-P_{b}dV+\mu dN$ are defined \citep[§3.1]{Katz2003-KATTOS}.\footnote{Although it has to be stressed that this relation is derived in the idealised situation described in \S4.1.} This relation derived within equilibrium statistical mechanics from the base micro-level description can be regarded as an alternative statement of the first law of thermodynamics, although this is not the general first law of thermodynamics in terms of work and heat introduced by phenomenological thermodynamics independently from any micro-description, i.e. $dU=\dbar Q + \dbar W$. This is thus a good example of a relation that is not fully thermodynamic in the sense of phenomenological thermodynamic descriptions but is also not based on micro-variables, and supports (I). Hence minimal thermodynamics captures a middle ground level of description between the two levels.

It is crucial to highlight the role of limits and how their use reinforces our ability to provide thermodynamic descriptions in SGS. As discussed in the previous section, we can meaningfully formulate most of these quantities and relations thanks to the application of a limit ($N \rightarrow \infty$) which is similar to the thermodynamic limit. The use of this type of thermodynamic limit is a further reason to regard this as a kind of thermodynamic level of description, as per point (I). This consideration also allows us to respond to one of Robertson's arguments reported in §3.1. Robertson claims that the inapplicability of thermodynamics to SGS is unsurprising in light of the failure of the thermodynamic limit in SGS. She also stresses that in the limit probabilistic descriptions become non-probabilistic, certain crucial statistical quantities become extensive, and ensembles are equivalent. We showed that, whereas the $N, V \rightarrow \infty$ (constant $N/V$) limit does not work, another thermodynamic limit which allows us to recover higher-level non-statistical descriptions exists. Hence, on Robertson's grounds, a thermodynamic description \textit{can} be applied to SGS. Furthermore, addressing Callender's worry that taking the $N \rightarrow \infty$ limit as the right thermodynamic limit for this context could be question-begging, we stress that redefining thermodynamic relations and equilibrium states in this limit is not a mere philosophical exercise, but rather allows us to develop useful thermodynamic description like the one involved in Katz's stability theory, and so this is a principled choice.

All these things being considered, this section has clarified how the theory of phenomenological thermodynamics is unsuitable for SGS. If we think of thermodynamics purely in terms of phenomenological thermodynamics \hl{as we introduced it} then it is reasonable to conclude that thermodynamics does not apply to SGS as Robertson concludes, \hl{by definition. Concepts such as thermal contact, work, heat, and the gradual variation of external parameters, along with operations performed on these systems, hold little relevance — or are entirely absent —from the actual physics literature on SGS}.\footnote{See \citet[p. 19]{heggie_hut_2003} on the lack of handles we have on modelling SGS.} \hl{On this, we agree with Robertson.} Instead, the kind of `thermodynamic' descriptions we can develop in the physics of SGS match more with the type of thermodynamic relations we can formulate in equilibrium statistical mechanics. \hl{These relations fall within the domain of minimal thermodynamic descriptions rather than phenomenological thermodynamic descriptions.} For instance, as we mentioned, the relation $dE = TdS-P_{b}dV+\mu dN$ that holds for certain SGS models is the specific version of the first law of thermodynamics that is obtained in equilibrium statistical mechanics. That relation can in principle be used to derive the first law of thermodynamics $dU=\dbar Q + \dbar W$ as found in phenomenological thermodynamics, but they are not the same identical law, as one is empirically formulated and the other is derived within equilibrium statistical mechanics and they are based on different quantities. \hl{In this sense, they belong to two different levels of descriptions and abstraction, as articulated above.}\footnote{\hl{On this we agree with} \cite{Franklin2018-FRAEWL} \hl{who hold that `higher-level' descriptions obtained from abstractions and change of variables imposed on `lower-level' descriptions can be regarded as novel and weakly emergent even if reduction is available.}} \hl{Whereas the level of description at which quantities like energy and temperature is available and employed within the physics of SGS, descriptions of SGS via heat exchange or work are absent from the physics literature on SGS (assuming, of course, that it is even meaningful to discuss concepts like heat exchange within isolated astrophysical systems).} Furthermore, focusing on $T$, we stress that this is the temperature defined via statistical equilibrium and not temperature as introduced by thermodynamic equilibrium in phenomenological thermodynamics via the Zeroth law of thermodynamics, \hl{bolstering the claim that these are two descriptive levels.}

Wrapping up, there is still space for thermodynamics in SGS physics. The minimal framework for thermodynamics presented here is the most perspicuous way to think about thermodynamics in SGS especially as it is grounded in physics practice. Also, describing minimal thermodynamics as a kind of thermodynamics vindicates the `thermodynamic' label used by physicists in talking about thermodynamic methods in statistical SGS physics.\footnote{For instance \cite{chavanis2006phase} distinguished between the thermodynamic approach used by \cite{lynden1968gravo}, \cite{padmanabhan1990statistical} and himself and the statistical mechanical approach starting by the density of states or partition functions.} More broadly, adopting this framework clarifies how kinetic theory, equilibrium statistical mechanics, and a kind of thermodynamics embedded within the theoretical frame of equilibrium statistical mechanics are all employed at different levels in the physics of SGS. Every one of these three levels is equally justified and interpreting the physics of SGS leaving out one of these dimensions of description and explanation would miss something about our understanding of these systems. 

\section{Conclusion and Future Directions}

We have shown that equilibrium statistical mechanics and a minimal notion of thermodynamics can be applied to self-gravitating systems. Our analysis accounts for scientific practice and improves our understanding of the physics of SGS. More broadly, it sheds new light on concepts such as equilibrium and advances a new useful conceptualisation of thermodynamics in between statistical mechanical and purely phenomenological thermodynamic descriptions. 

Looking at future applications, this minimal conception of thermodynamics can help us understand other areas of physics, such as the physics of \textit{stars}, which shares important similarities with the physics of SGS. Statistical physics can be employed to model stars, but features like stability and equilibrium work in peculiar ways. Another area in which statistical physics can be useful is \textit{dark matter}. For instance, \cite{chavanis2015darkmatter} model dark matter halos using the King model. These systems bear interesting similarities with the SGS studied here. Furthermore, further areas where long-range interactions are involved include two-dimensional or geophysical fluid dynamics and plasma models. \hl{Finally, one crucial area where the results of this paper may have significant implications is the debate on black hole thermodynamics. Black holes are self-gravitating systems, and in recent years a rich body of literature has emerged questioning whether black holes can be regarded as thermodynamic objects} (\cite{Dougherty}, \cite{Wallace2018-WALTCF-9}, \cite{prunkl2019black}). \hl{The debate especially focuses on the alleged analogy between the laws of black hole mechanics and the laws of phenomenological thermodynamics, though the discussion extends beyond that aspect. Black holes are sui generis entities and exhibit peculiar features that the classical SGS systems considered in this paper do not possess. Future work could apply our findings to better understand black hole thermodynamics, particularly in relation to our discussion of idealized and realistic SGS models and our proposal to introduce a minimal thermodynamic level of description. An interesting question to explore would be the extent to which black holes can be modelled using phenomenological thermodynamic descriptions, or whether they should instead be described in terms of minimal thermodynamics.}

\section*{Acknowledgements}

This paper has greatly benefited from extensive conversations with David Wallace. I'm also grateful to Karim Thebault and James Ladyman for their valuable feedback, and to Douglas Heggie, Scott Tremaine, and Pierre-Henri Chavanis for insightful discussions. Finally, thanks to the audiences in Milan and Pittsburgh for their helpful questions and comments on earlier versions of this work. Any remaining errors are entirely my own. This paper was supported by the Swiss National Science Foundation, the SNSF Starting Grant project `Temporal existence', project number TMSGI1-211294, and by the South, West and Wales doctoral training partnership.

\bibliography{library}

\end{document}